\documentclass{emulateapj}
\usepackage{apjfonts}

\usepackage{mathrsfs}
\usepackage{CJK}

\usepackage{amsmath}

\usepackage{graphics}

\slugcomment{Received 2015 July 29; accepted 2015 November 30}

\shorttitle{An accretion disk-outflow model for hysteretic state
transition in X-ray binaries}

\shortauthors{X. Cao}

\begin{document}

\title{An accretion disk-outflow model for hysteretic state
transition in X-ray binaries}

\author{Xinwu Cao\altaffilmark{1,2}}
\altaffiltext{1}{SHAO-XMU Joint Center for Astrophysics, Shanghai
Astronomical Observatory, Chinese Academy of Sciences, 80 Nandan
Road, Shanghai, 200030, China; cxw@shao.ac.cn} \altaffiltext{2}{Key
Laboratory of Radio Astronomy, Chinese Academy of Sciences, 210008
Nanjing, China}

\begin{abstract}
We suggest a model of the advection dominated accretion flow (ADAF)
with magnetically driven outflows to explain the hysteretic state
transition observed in X-ray binaries (XRBs). The transition from a
thin disk to an ADAF occurs when the mass accretion rate is below a
critical value. The critical mass accretion rate for the ADAF can be
estimated by equating the equilibration timescale to the accretion
timescale of the ADAF, which is sensitive to its radial velocity.
The radial velocity of thin disks is very small, which leads to the
advection of the external field in thin disks becoming very
inefficient. ADAFs are present in the low/hard states of XRBs, and
their radial velocity is large compared with the thin disk. The
external field can be dragged inward efficiently by the ADAF, so a
strong large-scale magnetic field threading the ADAF can be formed,
which may accelerate a fraction of gas in the ADAF into the
outflows. Such outflows may carry away a large amount of angular
momentum from the ADAF, which significantly increases the radial
velocity of the ADAF. This leads to a high critical mass accretion
rate, below which an ADAF with magnetic outflows can survive. Our
calculations show that the critical luminosity of the ADAF with
magnetic outflows can be one order of magnitude higher than that for
a conventional ADAF, if the ratio of gas to magnetic pressure
$\beta\sim 4$ in the disk. This can naturally explain the hysteretic
state transition observed in XRBs.
\end{abstract}

\keywords{accretion, accretion disks, black hole physics, magnetic
fields, X-rays: binaries}


\section{Introduction}\label{intro}

Variability of X-ray binaries (XRBs) is closely related to accretion
disks, which exhibits transitions between a thermal state, a
substantial thermal component in the continuum spectra, and a
low/hard state dominated by non-thermal emission. It is believed
that the states of X-ray binaries correspond to different accretion
modes. Hot accretion flows are present in low/hard state, while the
black holes are surrounded by geometrically thin accretion disks in
the thermal state of XRBs
\citep*[e.g.,][]{1997ApJ...489..865E,2000A&A...354L..67M,2004MNRAS.351..791Z,2008ApJ...682..212W,2010LNP...794...53B,2014ApJ...788...52C,2015MNRAS.453.3447D}.

The accretion mode transition is mainly triggered by the
dimensionless mass accretion rate $\dot{m}$
($\dot{m}=\dot{M}/\dot{M}_{\rm crit}$), which depends sensitively on
the viscosity parameter $\alpha$ \citep*[see][for
reviews]{1998tbha.conf..148N,2014ARA&A..52..529Y}. The critical mass
accretion rate $\dot{m}_{\rm crit}$ can be estimated by equating the
ion-electron equilibration timescale to the accretion timescale of
an advection dominated accretion flow (ADAF)
\citep*[][]{1998tbha.conf..148N}. Although the detailed physics of
the accretion mode transition is still unclear, it may be regulated
by the processes in the accretion disk corona systems, i.e., the
evaporation of the disk and/or condensation of the hot gas in the
corona \citep*[][]{1999A&A...348..154M,1999ApJ...527L..17L}.

However, the hysteretic state transition in XRBs, i.e., the
transition of the low/hard state to thermal state occurs at a
luminosity much higher than that for the transition from the thermal
state to low/hard state
\citep*[e.g.,][]{1995ApJ...442L..13M,1997ApJ...477L..95Z,2002MNRAS.332..856N,2003MNRAS.338..189M,2004ApJ...611L.121Y,2004MNRAS.351..791Z,2009ApJ...701.1940Y},
is still not well understood in the frame of the above mentioned
scenarios \citep*[see, e.g.,][for a review]{2013FrPhy...8..630Z}. In
the most of the previous models, the accretion mode transition is
solely triggered by the dimensionless mass accretion rate
\citep*[see, e.g.,][for
reviews]{1998tbha.conf..148N,2014ARA&A..52..529Y}. In order to solve
this problem, several different scenarios have been suggested
\citep*[e.g.,][]{2005A&A...432..181M,2008MNRAS.385L..88P,2008ApJ...674..408B,2014ApJ...782L..18B,2014MNRAS.437.3994N,2015arXiv150703996B}.
In the frame of the evaporation model, it was assumed that there are
different amounts of Compton cooling or heating acting on the disk
in different accretion modes, which leads to the hysteretic state
transition in XRBs
\citep*[][]{2005A&A...432..181M,2005A&A...442..555L}. Alternatively,
the hysteretic state transition is assumed to be related to magnetic
processes in the disks
\citep*[e.g.,][]{2008MNRAS.385L..88P,2008ApJ...674..408B,2014ApJ...782L..18B,2015arXiv150907156L}.
\citet{2008ApJ...674..408B} suggested that there are two different
regions with $P_{\rm m}>1$ or $P_{\rm m}<1$ ($P_{\rm m}$ is the
magnetic Prandtl number) in the accretion disk. The transition
radius of these two regions is regulated by the dimensionless
accretion rate $\dot{m}$, and the XRB state transition is related to
an unstable interface between these two regions in the disk. The
critical dimensionless mass accretion rate for the accretion mode
transition depends sensitively on the radial velocity of the
accretion flow, which is a function of $\alpha$. Motivated by
numerical simulations, the hysteretic state transition can be
explained if the value of $\alpha$ is assumed to vary in different
types of accretion disks, i.e., $\alpha\sim 1$ in ADAFs, while
$\alpha\sim 0.01-0.1$ in standard thin disks
\citep*[][]{2014ApJ...782L..18B}.

Magnetic fields play an important role in some of these scenarios,
though the origin the fields is still unclear. One possibility is
that the external generated large-scale poloidal field is dragged
inward by the plasma in the disk, while the field diffuses outward
simultaneously
\citep*[][]{1974Ap&SS..28...45B,1976Ap&SS..42..401B,1989ASSL..156...99V,1994MNRAS.267..235L,2001ApJ...553..158O}.
However, the advection of the field in a geometrically thin ($H/R\ll
1$) is inefficient due to its small radial velocity. The magnetic
diffusion timescale is about the same as the viscous timescale in a
steady disk, which leads to a very weak radial field component at
the surface of a thin disk \citep*[][]{1994MNRAS.267..235L}, though
some specific mechanisms were suggested to alleviate the difficulty
of field advection in thin disks
\citep*[][]{2005ApJ...629..960S,2009ApJ...701..885L,2012MNRAS.424.2097G,2013MNRAS.430..822G,2013ApJ...765..149C}.
The vertical field can be dragged efficiently by an ADAF
\citep*[][]{2011ApJ...737...94C}, because the radial velocity of
ADAFs is much larger than that of a thin accretion disk
\citep*[][]{1994ApJ...428L..13N,1995ApJ...452..710N}.

In this work, we suggest that outflows are driven by the magnetic
field advected by the ADAF, which carry a large amount of angular
momentum away from the ADAF. The radial velocity of the ADAF with
magnetically driven outflows can be significantly higher than that
for a conventional ADAF without magnetic outflows. This leads to a
higher critical mass accretion rate of the transition from an ADAF
to a thin disk. In Sect. \ref{mdot_crit_adaf}, we estimate the
critical accretion rate of the transition from an ADAF with magnetic
outflows to a thin disk. The properties of such magnetically driven
outflows are analyzed in Sect. \ref{outflow_b}. Section 4 contains
the discussion of the model. The last section is a brief summary.

\vskip 1cm
\section{The critical mass accretion rate for ADAFs with magnetically driven outflows}\label{mdot_crit_adaf}

There is a critical mass accretion rate $\dot{m}_{\rm crit}$, above
which the hot ADAF is switched to an optically thick accretion disk
\citep*[][]{1997ApJ...489..865E}. The hot ADAF is a two-temperature
accretion flow, in which the ion temperature is significantly higher
than the electron temperature. The energy of the ions is transported
to electrons via Coulomb collisions
\citep*[][]{1983MNRAS.204.1269S}. The critical mass accretion rate
$\dot{m}_{\rm crit}$ can be estimated by equating an ion-electron
equilibration timescale to an accretion timescale of an ADAF
\citep*[][]{1998tbha.conf..148N}. The ion-electron equilibration
timescale,
\begin{equation}
t_{\rm ie}\sim{\frac {u}{q_{\rm ie}}},\label{t_ie}
\end{equation}
where the internal energy of the gas is
\begin{equation}
u={\frac {3}{2}}n_{\rm i}kT_{\rm i}+{\frac {3}{2}}n_{\rm e}kT_{\rm
e}\simeq {\frac {3}{2}}n_{\rm i}kT_{\rm i}, \label{u}
\end{equation}
because the internal energy of the elections is negligible in ADAFs.
The Coulomb interaction between the electrons and ions is given by
\citep*[][]{1983MNRAS.204.1269S,1995ApJ...452..710N,1998MNRAS.296L..51Z}
\begin{displaymath}
q_{\rm ie}={\frac {3}{2}}{\frac {m_{\rm e}}{m_{\rm p}}}n_{\rm
e}n_{\rm i}\sigma_{\rm T}c{\frac {kT_{\rm i}-kT_{\rm
e}}{K_2(1/\Theta_{\rm e})K_2(1/\Theta_{\rm
i})}}\ln\Lambda~~~~~~~~~~~~~~~~~~~~~~~~~~
\end{displaymath}
\begin{equation}
\times \left[{\frac {2(\Theta_{\rm i}+\Theta_{\rm
e})^2+1}{\Theta_{\rm i}+\Theta_{\rm e}}}K_1\left({\frac {\Theta_{\rm
i}+\Theta_{\rm e}}{\Theta_{\rm i}\Theta_{\rm
e}}}\right)+2K_0\left({\frac {\Theta_{\rm i}+\Theta_{\rm
e}}{\Theta_{\rm i}\Theta_{\rm e}}}\right)\right],\label{q_ie}
\end{equation}
where $\Theta_{\rm i}=kT_{\rm i}/m_{\rm p}c^2$, $\Theta_{\rm
e}=kT_{\rm e}/m_{\rm e}c^2$, and $\ln\Lambda=20$ is adopted. We
rewrite Eq. (\ref{t_ie}) as
\begin{equation}
t_{\rm ie}\sim{\frac {u}{q_{\rm ie}}}\simeq {\frac {f(\Theta_{\rm
i},\Theta_{\rm e})}{n_{\rm e}\sigma_{\rm T}c}},\label{t_ie2}
\end{equation}
where
\begin{displaymath}
f(\Theta_{\rm i},\Theta_{\rm e})={\frac {m_{\rm p}}{m_{\rm
e}}}{\frac {T_{\rm i}}{T_{\rm i}-T_{\rm e}}}{\frac
{K_2(1/\Theta_{\rm e})K_2(1/\Theta_{\rm
i})}{\ln\Lambda}}~~~~~~~~~~~~~~~~~~~~~~~~~~~~~~~
\end{displaymath}
\begin{equation}
\times  \left[{\frac {2(\Theta_{\rm i}+\Theta_{\rm
e})^2+1}{\Theta_{\rm i}+\Theta_{\rm e}}}K_1\left({\frac {\Theta_{\rm
i}+\Theta_{\rm e}}{\Theta_{\rm i}\Theta_{\rm
e}}}\right)+2K_0\left({\frac {\Theta_{\rm i}+\Theta_{\rm
e}}{\Theta_{\rm i}\Theta_{\rm
e}}}\right)\right]^{-1}.\label{f_theta}
\end{equation}
The accretion timescale is
\begin{equation}
t_{\rm acc}\sim {\frac {R}{|V_{R}|}}={\frac {R^2}{\alpha\Omega_{\rm
K}H^2}}, \label{t_acc}
\end{equation}
in which the radial velocity of the ADAF,
\begin{equation}
V_R\simeq -\alpha c_{\rm s}{\frac {H}{R}}=-\alpha\Omega_{\rm
K}{\frac {H^2}{R}}, \label{v_r}
\end{equation}
has been adopted.

The ADAF is suppressed when it is accreting at
$\dot{M}\ga\dot{M}_{\rm crit}$. The value of $\dot{M}_{\rm crit}$
can be estimated with $t_{\rm ie}=t_{\rm acc}$, i.e.,
\begin{equation}
{\frac {f(\Theta_{\rm i},\Theta_{\rm e})}{n_{\rm e}\sigma_{\rm
T}c}}={\frac {R^2}{\alpha\Omega_{\rm K}H^2}}. \label{t_iet_acc}
\end{equation}
With Eq. (\ref{t_iet_acc}), we can estimate the critical mass
accretion rate as
\begin{equation}
\dot{M}_{\rm crit}=-2\pi R(2H\rho)V_R=\alpha^2\dot{M}_{\rm crit,0},
\label{mdot_crit}
\end{equation}
where
\begin{equation}
\dot{M}_{\rm crit,0}=4\pi m_{\rm p}\left({\frac
{H}{R}}\right)^2{\frac {H^3\Omega_{\rm K}^2}{\sigma_{\rm
T}c}}f(\Theta_{\rm i},\Theta_{\rm e}) \label{mdot_crit2}
\end{equation}
is the critical mass accretion rate corresponding to the case of
$\alpha=1$.

In a similar way, we consider an ADAF with magnetically driven
outflows carrying most of the angular momentum away from the
accretion flow. The structure of the disk will be altered in the
presence of magnetic outflows
\citep*[][]{2009MNRAS.400.1734L,2013ApJ...765..149C,2014ApJ...786....6L,2014ApJ...788...71L}.
{The angular momentum equation for a steady ADAF with outflows is
\begin{equation}
{\frac {d}{dR}}(2\pi R \Sigma V_R^\prime R^2\Omega)={\frac
{d}{dR}}(2\pi R \nu \Sigma R^2{\frac {d\Omega}{dR}})-2\pi RT_{\rm
m}, \label{angular_1}
\end{equation}
where $T_{\rm m}$ is the magnetic torque due to the outflows (we use
``$\prime$" to describe the quantities of the ADAF with outflows
hereafter). Integrating Equation (\ref{angular_1}), yields
\begin{equation}
2\pi R\Sigma V_R^\prime R^2\Omega=2\pi R\nu\Sigma R^2{\frac
{d\Omega}{dR}}-2\pi\int RT_{\rm m}(R)dR+C, \label{angular_2}
\end{equation}
where the value of the integral constant $C$ can be determined with
a boundary condition on the accreting object. The magnetic torque
$T_{\rm m}(R)$ can be calculated if the global outflow solution is
derived, which is beyond the scope of this work. For simplicity, we
assume $T_{\rm m}\propto R^{-\xi_{\rm T}}$, Eq. (\ref{angular_2})
becomes
\begin{equation}
V_R^\prime\simeq -{\frac {\nu}{R}}-{\frac {1}{2-\xi_{\rm T}}}{\frac
{T_{\rm m}}{R\Sigma\Omega}}, \label{v_r0}
\end{equation}
where the approximations, $d\Omega/dR\simeq -\Omega/R$ is adopted,
and $C$ is negligible in the region not very close to the inner disk
edge \citep*[][]{1973A&A....24..337S,1994ApJ...428L..13N}. In the
self-similar ADAF solution, the gas pressure $p_{\rm gas}\propto
R^{-5/2}$ \citep*[][]{1995ApJ...452..710N}. Assuming the magnetic
pressure $B^2\propto p_{\rm gas}$, we have $B\propto R^{-5/4}$, and
$T_{\rm m}\propto RB^2\propto R^{-3/2}$. Thus, we adopt $\xi_{\rm
T}=3/2$ in all of our calculations. In this work, we use a model
parameter $f_{\rm m}$ to describe the relative importance of the
angular momentum removal of the disk, so the main conclusions of
this work will not be altered by the precise value of $\xi_{\rm T}$
adopted.} In this case, the radial velocity of the ADAF is
\begin{equation}
V_R^\prime\simeq V_{R}+V_{R,\rm m},\label{v_r2}
\end{equation}
where the first term is due to the conventional turbulence in the
disk, and the second term is contributed by the outflows,
\begin{equation}
V_{R,\rm m}=-{\frac {2T_{\rm m}}{R\Sigma\Omega}}.\label{v_rm}
\end{equation}
We use a parameter $f_{\rm m}$ to describe the relative importance
of the outflows on the radial velocity of the ADAF,
\begin{equation}
V_R^\prime=(1+f_{\rm m})V_{R}=-(1+f_{\rm m})\alpha\Omega_{\rm
K}{\frac {H^2}{R}}.\label{v_r3}
\end{equation}
The value of $f_{\rm m}$ is determined by the properties of the ADAF
and the magnetically driven outflows, which will be discussed in
Sect. \ref{outflow_b}. The accretion timescale for the ADAF with
magnetically driven outflows is
\begin{equation}
t_{\rm acc}^\prime\sim {\frac {R}{|V_{R}^\prime|}}={\frac
{R^2}{\alpha\Omega_{\rm K}H^2(1+f_{\rm m})}}, \label{t_acc2}
\end{equation}
In a similar way as the conventional ADAF case discussed above, the
value of $\dot{M}_{\rm crit}^\prime$ can be estimated with $t_{\rm
ie}^\prime=t_{\rm acc}^\prime$. {Almost all gravitational energy is
carried by the ions in the accretion disk, and the ions are heated
to high temperatures if the energy transfer from the ions to
electrons is inefficient. The ion temperature is therefore nearly
virialized in ADAFs \citep*[see, e.g.,][for a
review]{2014ARA&A..52..529Y}. We believe that it is also the case
for the ADAF with outflows accreting at the critical rate, i.e.,
$\Theta_{\rm i}^\prime\simeq \Theta_{\rm i}$, because the
ion-electron equilibration timescale is comparable to the accretion
timescale for the ADAF with outflows, which is the same as the
conventional ADAF case.} The ion-electron equilibration timescale
$t_{\rm ie}^\prime$ only depends on the local properties of the gas
in the ADAF, which can be estimated by the same equation
(\ref{t_ie2}) for the conventional ADAF. Thus, we have
\begin{equation}
{\frac {f(\Theta_{\rm i},\Theta_{\rm e}^\prime)}{n_{\rm
e}^\prime\sigma_{\rm T}c}}={\frac {R^2}{\alpha\Omega_{\rm
K}{H^2(1+f_{\rm m})}}}, \label{t_iet_acc2}
\end{equation}
and the critical mass accretion rate for the ADAF with magnetic
outflows is estimated as
\begin{equation}
\dot{M}_{\rm crit}^\prime=-2\pi
R(2H^\prime\rho^\prime)V_R^\prime=\alpha^2(1+f_{\rm m})^2{\frac
{f(\Theta_{\rm i},\Theta_{\rm e}^\prime)}{f(\Theta_{\rm
i},\Theta_{\rm e})}}\dot{M}_{\rm crit,0}, \label{mdot_crit3}
\end{equation}
where Eqs. (\ref{v_r3}) and (\ref{t_iet_acc2}) have been used.

The gravitational energy dissipation rate from the unit area of the
disk is
\begin{equation}
Q_+={\frac {1}{2}}\nu\Sigma\left(R{\frac {d\Omega}{dR}}\right)^2,
\label{q_plus}
\end{equation}
where $\nu=\alpha c_{\rm s}H$. Substituting Eqs. (\ref{v_r}) and
(\ref{mdot_crit}) into Eq. (\ref{q_plus}), we derive the energy
dissipation rate of a conventional ADAF accreting at the critical
rate as
\begin{equation}
Q_+={\frac {1}{4\pi}}\left(R{\frac
{d\Omega}{dR}}\right)^2\dot{M}_{\rm crit}={\frac
{1}{4\pi}}\left(R{\frac {d\Omega}{dR}}\right)^2\alpha^2\dot{M}_{\rm
crit,0}. \label{q_plus2}
\end{equation}
For the ADAF with magnetically driven outflows, we substitute Eqs.
(\ref{v_r3}) and (\ref{mdot_crit3}) into Eq. (\ref{q_plus}), and the
gravitational energy dissipation rate is available,
\begin{displaymath}
Q_+^\prime={\frac {1}{4\pi}}\left(R{\frac
{d\Omega}{dR}}\right)^2{\frac {\dot{M}_{\rm crit}^\prime}{(1+f_{\rm
m})}}~~~~~~~~~~~~~~~~~~~~~~~~~~~~~~~~~~~~~~~~~~~~~~~~~
\end{displaymath}
\begin{equation}
={\frac {1}{4\pi}}\left(R{\frac {d\Omega}{dR}}\right)^2{\frac
{f(\Theta_{\rm i},\Theta_{\rm e}^\prime)}{f(\Theta_{\rm
i},\Theta_{\rm e})}}(1+f_{\rm m})\alpha^2\dot{M}_{\rm
crit,0}.~~~~~~~~ \label{q_plus3}
\end{equation}
For an ADAF accreting at the critical rates, the luminosity is
\begin{equation}
L_{\rm ADAF}\sim\int 2\pi RQ_+dR, \label{l_adaf}
\end{equation}
since the advection is not important for the ADAFs accreting at a
critical rate
\citep*[e.g.,][]{1998tbha.conf..148N,2014ARA&A..52..529Y}.

{The local density of an ADAF with magnetic outflows accreting at
the critical rate is the same as that of a conventional ADAF
accreting at $(1+f_{\rm m})^2\dot{M}f(\Theta_{\rm i},\Theta_{\rm
e}^\prime)/f(\Theta_{\rm i},\Theta_{\rm e})$ with a large viscosity
parameter of $(1+f_{\rm m})\alpha$, while the energy dissipation
rate of the ADAF with magnetic outflows is only $\sim[(1+f_{\rm
m})f(\Theta_{\rm i},\Theta_{\rm e}^\prime)/f(\Theta_{\rm
i},\Theta_{\rm e})]^{-1}$ times of this conventional ADAF. It means
that the electron temperature of the ADAF with outflows is lower
than that of this conventional ADAF, i.e., $\Theta_{\rm
e}^\prime<\Theta_{\rm e}$, which leads to $f(\Theta_{\rm i}^\prime,
\Theta_{\rm e}^\prime)>f(\Theta_{\rm i}, \Theta_{\rm e})$
($\Theta_{\rm i}^\prime\sim \Theta_{\rm i}$). In this work, we
conservatively adopt the lower limits of $M_{\rm crit}^\prime$ and
$Q_+^\prime$ for the ADAF with outflows in our discussion, i.e.,
\begin{equation}
\dot{M}_{\rm crit}^\prime=-2\pi
R(2H\rho)V_R^\prime\simeq\alpha^2(1+f_{\rm m})^2\dot{M}_{\rm
crit,0}, \label{mdot_crit4}
\end{equation}
and
\begin{equation}
Q_+^\prime={\frac {1}{4\pi}}\left(R{\frac
{d\Omega}{dR}}\right)^2{\frac {\dot{M}_{\rm crit}^\prime}{(1+f_{\rm
m})}}\simeq{\frac {1}{4\pi}}\left(R{\frac
{d\Omega}{dR}}\right)^2(1+f_{\rm m})\alpha^2\dot{M}_{\rm crit,0}.
\label{q_plus4}
\end{equation}}This means that $L_{\rm ADAF}^\prime\sim (1+f_{\rm m})L_{\rm ADAF}$,
i.e., the ADAF with magnetic outflows is $\sim f_{\rm m}$ times more
luminous than the conventional ADAF accreting at the critical rate.
The critical mass accretion rate for the ADAF with magnetic outflows
is $\sim (1+f_{\rm m})^2$ times of that for a conventional ADAF.
{Only $\sim 1/(1+f_{\rm m})$ of the kinetic energy is dissipated in
the accretion flow. The luminosity of the ADAF with magnetic
outflows is not proportional to its mass accretion rate because
$\sim f_{\rm m}/(1+f_{\rm m})$ of the kinetic energy of the disk is
tapped to accelerate the outflows.}

{The cooling timescale of an ADAF is estimated by
\begin{equation}
t_{\rm cool}\sim {\frac {2Hu}{F_{\rm rad}}}, \label{t_cool}
\end{equation}
where $F_{\rm rad}$ is the radiation flux from the unit surface area
of the disk. The radiation flux $F_{\rm rad}$ is related to the
gravitational energy rate $Q_+$ by
\begin{equation}
F_{\rm rad}=(1-f_{\rm adv})Q_+, \label{f_rad}
\end{equation}
where $f_{\rm adv}$ is the fraction of the dissipated energy
advected in the flow. Equation (\ref{t_cool}) can be rewritten as
\begin{equation}
t_{\rm cool}\sim {\frac {2Hu}{(1-f_{\rm adv})Q_+}}={\frac
{3}{\alpha\Omega_{\rm K}f_{\Omega}^2(1-f_{\rm adv})}},
\label{t_cool2}
\end{equation}
where $c_{\rm s}=H\Omega_{\rm K}$, $f_\Omega=\Omega/\Omega_{\rm K}$,
Eq. (\ref{u}), and the approximation
\begin{equation}
Q_+={\frac {1}{2}}\nu\Sigma\left(R{\frac
{d\Omega}{dR}}\right)^2\simeq {\frac {1}{2}}\alpha c_{\rm
s}H\Sigma\Omega^2 \label{q_plus4}
\end{equation}
are used. We note that the cooling timescale of the flow only
depends on the local properties of the ADAF, which is not explicitly
relevant to the magnetic outflows. This means that Eq.
(\ref{t_cool2}) is valid either for a conventional ADAF or an ADAF
with magnetic outflows.}

{The advection of the external magnetic field is balanced by the
magnetic diffusion in a steady accretion flow, and therefore the
magnetic diffusion timescale is comparable to the accretion
timescale of the flow, i.e., $t_{\rm dif}\sim t_{\rm acc}$
\citep*[][]{1994MNRAS.267..235L,2013ApJ...765..149C}. For the
magnetic field of an ADAF with magnetic outflows, its magnetic
diffusion timescale is
\begin{equation}
t_{\rm dif}^\prime\sim t_{\rm acc}^\prime\sim {\frac
{R}{|V_{R}^\prime|}}={\frac {R^2}{\alpha\Omega_{\rm K}H^2(1+f_{\rm
m})}}. \label{t_dif}
\end{equation}
The duration of the transition from an ADAF to a thin disk is at the
order of the cooling timescale of the ADAF (see Eq. \ref{t_cool2}).
After such an accretion mode transition, the strong magnetic field
of the ADAF decays at the magnetic diffusion timescale (see Eq.
\ref{t_dif}). For ADAFs accreting at critical rates, $H/R\sim 1$,
$f_\Omega\la 1$, and $f_{\rm adv}\sim 0$
\citep*[][]{2014ARA&A..52..529Y}, the magnetic diffusion timescale,
\begin{equation}
t_{\rm dif,ADAF}^\prime=f_{\Omega}^2(1-f_{\rm adv})\left({\frac
{R}{H}}\right)^2{\frac {t_{\rm cool}^\prime}{3(1+f_{\rm m})}}\sim
{\frac {t_{\rm cool}^\prime}{3(1+f_{\rm m})}}, \label{t_dif2}
\end{equation}
is much shorter than the cooling timescale. {The magnetic diffusion
timescale of a thin/slim disk is
\begin{equation}
t_{\rm dif,SD}\sim {\frac {RH_{\rm SD}\kappa_0}{\eta}}={\frac
{R\kappa_0}{\alpha c_{\rm s}P_{\rm m}}}, \label{t_dif3}
\end{equation}
where $H_{\rm SD}$ is the half-thickness of the disk,
$\kappa_0=B_z/B_{R,{\rm s}}$ is the inclination of the field line at
the disk surface, $\eta$ is the magnetic diffusivity, and the
magnetic Prandtl number is $P_{\rm m}=\eta/\nu=\eta/\alpha c_{\rm
s}H$. Compared to the cooling timescale of the ADAF with outflows
accreting at the critical rate, we have
\begin{equation}
t_{\rm dif,SD}\sim{\frac {1}{3}}\left({\frac {H_{\rm SD}}{R}}
\right)^{-1}{\frac {\kappa_0}{P_{\rm m}}}t_{\rm cool}^\prime.
\label{t_dif4}
\end{equation}
The critical mass accretion rate can be close to the Eddington rate
provided $\alpha\sim 0.1$ and $f_{\rm m}\sim 10$ (see Equation
\ref{mdot_crit4}), and therefore the ADAF with outflows may transit
to a slim disk. The relative half-thickness $H_{\rm SD}/R\sim 0.5$
for a slim disk accreting at the Eddington rate \citep*[see,
e.g.,][]{2010ApJ...715..623L}. For the typical values of the
parameters, $P_{\rm m}\sim 1$ and $\kappa_0\sim 1$ \citep*[see][for
a detailed discussion, and the references
therein]{2013ApJ...765..149C}, we estimate the magnetic diffusion
timescale of a slim disk accreting at the Eddington rate, $t_{\rm
dif,SD}\sim t_{\rm cool}^\prime$ (see Equation \ref{t_dif4}). }
%
}

\vskip 1cm

\section{The magnetically driven outflows}\label{outflow_b}

For an ADAF with magnetically driven outflows, the magnetic torque
exerted by the outflows in unit area of the disk surface is
\begin{equation}
T_{\rm m}={\frac {B_zB_{\phi}^{\rm s}}{2\pi}}R,\label{t_m_1}
\end{equation}
where $B_{\phi}^{\rm s}$ is the azimuthal component of the
large-scale magnetic field at the disk surface. The radial velocity
of the ADAF is
\begin{displaymath}
V_R^\prime=V_{R}+V_{R,\rm m}=-\alpha c_{\rm s}{\frac {H}{R}}-{\frac
{2T_{\rm m}}{\Sigma
R\Omega}}~~~~~~~~~~~~~~~~~~~~~~~~~~~~~~~~~~~~~~~~~~~~~~~~
\end{displaymath}
\begin{equation}
=-\alpha c_{\rm s}{\frac {H}{R}}- {\frac {B_zB_{\phi}^{\rm
s}}{\pi\Sigma\Omega}}=V_R\left(1+{\frac {B_zB_{\phi}^{\rm
s}}{\pi\Sigma\Omega}}{\frac {R}{\alpha c_{\rm
s}H}}\right),~~~~~~~~~~~~~\label{v_r4}
\end{equation}
which can be rewritten as
\begin{equation}
V_R^\prime=V_R\left(1+{\frac {4\xi_\phi}{\tilde{H}\beta\alpha
f_\Omega}}\right),\label{v_r5}
\end{equation}
where the dimensionless quantities are defined as
$\xi_\phi=-B_{\phi}^{\rm s}/B_z$, $\beta=8\pi p_{\rm gas}/B_z^2$,
and $\tilde{H}=H/R$. Comparison of Equations (\ref{v_r3}) with
(\ref{v_r5}) leads to
\begin{equation}
f_{\rm m}={\frac {4\xi_\phi}{\tilde{H}\beta\alpha f_\Omega}}.
\label{f_m}
\end{equation}
The ratio $\xi_\phi\la 1$ \citep*[see the detailed discussion
in][]{1999ApJ...512..100L}. For ADAFs, $\tilde{H}\sim 1$ and
$f_\Omega\la 1$, so we can estimate the value of $f_{\rm m}$ as
$f_{\rm m}\sim 4/\beta\alpha$. For a typical value of $\alpha=0.1$,
$f_{\rm m}$ can be as large as $\sim 10$ with magnetic field
strength of $\beta\sim 4$.

\vskip 1cm

\section{Discussion}\label{discussion}

It is known that the radial velocity of the disk $V_R\propto
(H/R)^2$, and therefore $V_R$ is very small for thin disks. {The
field advection in the thin accretion disk is quite inefficient, and
therefore the strength field threading the disk is very weak
\citep*[][]{1994MNRAS.267..235L}. When the mass accretion rate
decreases to a rate below $\sim\dot{M}_{\rm crit}$, the thin disk
transits to an ADAF. The ADAF is hot, and its radial velocity is
large compared to the thin disk. In the initial ADAF state, the
magnetic field of the ADAF is very weak, which is similar to the
thin disk. Due to the large radial velocity of the ADAF, the weak
external field can be dragged inward efficiently to form a strong
field at the order of accretion timescale
\citep*[][]{2011ApJ...737...94C}, which may drive a fraction of gas
from the ADAF to form outflows.} Such outflows may carry away a
large amount of angular momentum from the ADAF, which increases the
radial velocity of the ADAF significantly. The critical mass
accretion rate for the ADAF with magnetic outflows is estimated by
equating the equilibration timescale to the accretion timescale of
the ADAF \citep*[][]{1998tbha.conf..148N}. The accretion timescale
of the ADAF with magnetic outflows is much lower than that of a
conventional ADAF due to its large radial velocity. This leads to a
high critical mass accretion rate, below which an ADAF with magnetic
outflows can survive (see Sect. \ref{mdot_crit_adaf} for the
detailed calculations).

{Due to the mass loss in the outflows, the mass accretion rate of
the ADAF decreases with decreasing radius $R$. The mass-loss rate in
the outflows is regulated by the magnetic field
configuration/strength and the disk properties (e.g., the density
and temperature of the gas in the ADAF). The present analysis on the
critical accretion rate does not depend on the detailed properties
of the outflows (e.g., the mass-loss rate in the outflows). In this
work, we focus on the local disk properties, and the mass accretion
rate measured at a certain radius is considered in our model. The
properties of the outflow can be derived with the magnetic outflow
solution if suitable boundary conditions are provided
\citep*[e.g.,][]{1994A&A...287...80C,2014ApJ...783...51C}, which is
beyond the scope of this work. }

{In the low/hard state of the XRB, the ADAF with outflows will be
suppressed when the mass accretion rate increases above $\sim
\dot{M}_{\rm crit}^\prime$. The duration of the transition of an
ADAF to a thin disk is at the order of the cooling timescale $t_{\rm
cool}^\prime$ of the ADAF (see Eq. \ref{t_cool2}). In the accretion
mode transition, the magnetic field will decay at the magnetic
diffusion timescale $t_{\rm dif}^\prime$ (see Eq. \ref{t_dif}). We
find that $t_{\rm dif}^\prime$ is always much shorter than $t_{\rm
cool}^\prime$ of the ADAFs accreting at the critical rates (see Eq.
\ref{t_dif2}), {while the diffusion timescale of the slim disk
transited from the ADAF with outflows is comparable with the cooling
timescale $t_{\rm cool}^\prime$.} This means that the strong field
of the ADAF diffuses away rapidly during its transition to a
thin/slim disk, which implies that the field of the thin/slim disk
is too weak to accelerate outflows, though some mechanisms have been
suggested to solve strong field formation difficulty in thin disks
\citep*[][]{2005ApJ...629..960S,2009ApJ...701..885L,2012MNRAS.424.2097G,2013MNRAS.430..822G,2013ApJ...765..149C}.}

Our calculations show that the critical luminosity of the ADAF with
magnetic outflows can be one order of magnitude higher than that for
a conventional ADAF, if the ratio of gas to magnetic pressure
$\beta\sim 4$ in the disk for $\alpha=0.1$ (see Sect.
\ref{outflow_b} for the detailed analyses). This is roughly
consistent with the strength of the field advected in the inner
region of a conventional ADAF \citep*[][]{2011ApJ...737...94C}. The
magnetic field advected in the ADAF could be stronger, if the
angular momentum carried by the outflows is properly considered,
which leads to higher radial velocity. This is also consistent with
relativistic jets always being associated with the low/hard state in
X-ray binaries, while jets are switched off in the thermal state
\citep*[][]{1999ApJ...519L.165F,2001ApJ...554...43C,2003MNRAS.343L..99F,2003MNRAS.344...60G,2004ARA&A..42..317F,2004MNRAS.355.1105F},
because jets are believed to be related to strong magnetic fields
\citep*[][]{1977MNRAS.179..433B,1982MNRAS.199..883B}. The viscosity
parameter $\alpha=0.1$ corresponds to $\dot{m}_{\rm crit}\sim 0.01$,
which is roughly consistent with those derived from the transition
from the thermal state to the low/hard state in some XRBs.

{Our results can be compared to those of the MHD simulations on
accretion disks
\citep*[e.g.,][]{2003ApJ...592.1042I,2009ApJ...707..428B,2010MNRAS.408..752P,2011MNRAS.418L..79T,2012MNRAS.423.3083M,2012MNRAS.426.3241N}.
The magnetic pressure can dominate over the gas pressure in the
inner region of the disk, and it becomes a magnetic arrested disk
\citep*[][]{2003ApJ...592.1042I}. For a given external field, the
ADAF is magnetically arrested only if the accretion rate is lower
than a certain rate \citep*[][]{2011ApJ...737...94C}. In our model,
the ADAF with outflows is accreting at the critical rate, and it may
not be in magnetic arrested state. The field strength of the ADAF
with outflows considered in this work is much lower than that of the
magnetic arrested case. The gas to magnetic pressure ratio $\beta\ga
4$ required in our analysis should be reasonable. The MHD
simulations of magnetized ADAFs show that a large fraction of gas
goes into the outflows, and the radial velocity of such ADAFs can be
more than one order of magnitude higher than that of a self-similar
ADAF \citep*[see][for the details]{2012MNRAS.426.3241N}. In order to
explain the observed hysteretic state transitions in XRBs, the
radial velocity of an ADAF with outflows being up to $\sim 10$ times
higher than that of a conventional ADAF is required in some extreme
state transition cases. This is consistent with the numerical
simulation in \citet{2012MNRAS.426.3241N}. }

{The radial velocity of a conventional ADAF is
\begin{equation}
V_R\simeq -c_1\alpha V_{\rm K}, \label{adaf_vr}
\end{equation}
where $V_{\rm K}$ is the Keplerian velocity at $R$, and $c_1$ is a
function of the disk parameters \citep*[][]{1995ApJ...452..710N}.
Our calculation indicates that $c_1$ is in the range of $\sim
0.4-0.6$. If a typical $\alpha=0.1$ is adopted, the radial velocity
of the ADAF with magnetic outflows can be as high as about a half of
the free-fall velocity when $(1+f_{\rm m})\sim 10$. The critical
accretion rate will be close to the Eddington rate. The ADAF with
such extreme properties needs to be verified by the global ADAF
solution with magnetic outflows in the future.

\vskip 1cm

\section{Summary}

In the thermal state, the XRB contains a thin disk. The thin disk
transits to an ADAF perhaps due to the evaporation of the disk
\citep*[][]{1999A&A...348..154M,1999ApJ...527L..17L}, when the mass
accretion rate declines below the critical rate. In the low/hard
state, the external weak field of the gas from the companion star
can be dragged inward efficiently in the ADAF, which may drive
outflows and increase the radial velocity of the ADAF significantly.
Thus, the ADAF with magnetic outflows can survive at a higher
accretion rate than a conventional ADAF. With an increasing mass
accretion rate, the transition from an ADAF to a thin disk occurs at
a higher luminosity than that from the thin disk to an ADAF. The
ADAF-outflow model suggested in this work can naturally explain the
hysteretic state transition observed in XRBs.\\

~~~~~\\

\acknowledgments I thank the referee for the very insightful
comments. This work is supported by the NSFC (grants 11173043, and
11233006), the Strategic Priority Research Program ¡°the Emergence
of Cosmological Structures¡± of the CAS (grant No. XDB09000000), the
CAS/SAFEA International Partnership Program for Creative Research
Teams, and Shanghai Municipality.

{}

\end{document}